\documentclass[sigconf]{acmart}
\AtBeginDocument{%
  }

\copyrightyear{2026}
\acmYear{2026}
\setcopyright{cc}
\setcctype{by}
\acmConference[UIST Adjunct '26]{The 39th Annual ACM Symposium on User Interface Software and Technology}{November 02--05, 2026}{Detroit, MI, USA}
\acmBooktitle{The 39th Annual ACM Symposium on User Interface Software and Technology (UIST Adjunct '26), November 02--05, 2026, Detroit, MI, USA}
\acmDOI{10.1145/3830397.3841886}
\acmISBN{979-8-4007-2855-6/2026/11}

\begin{document}

\title{The Systems Paper is Dead. Long Live the Systems Paper.}

\author{Björn Hartmann}
\email{bjoern@eecs.berkeley.edu}
\orcid{0002-0693-0829}
\correspondingauthor
\affiliation{%
  \institution{UC Berkeley EECS}
  \city{Berkeley}
  \state{California}
  \country{USA}
}

\renewcommand{\shortauthors}{Hartmann}

\begin{abstract}
The way we structure, conduct, and write up interactive systems research in UIST papers rests on assumptions about constraints that may no longer hold today. \textit{What should an impactful UIST paper look like when building working systems is no longer hard?} I argue that it should look different, and that we should ask more of our papers once implementation stops being a bottleneck. 

\end{abstract}

\ccsdesc[500]{Human-centered computing~HCI theory, concepts and models}
\ccsdesc[500]{Human-centered computing~HCI design and evaluation methods}



\maketitle

\section{The UIST Paper Formula, and Why It May Be Time for Methodological Change}
Recent UIST visions have focused on how we should change \textit{what we build} in a time of rapid AI progress. Igarashi~\cite{10.1145/3746058.3762828} and Satyanarayan~\cite{10.1145/3672539.3695751} have argued that generative AI is changing the kinds of systems we should design,  identifying new goals once ``easy and fast'' content creation is solved, and focusing on dynamic reconfigurations of agency over static design of UI affordances. I'd like us to consider a higher-level question: do we as researchers need to change \textit{how we generate knowledge through building systems}?

The canonical UIST systems research process, described well by Hudson and Mankoff~\cite{Hudson2014}, goes like this: invent a clever new concept that bridges emerging technology and human need, then build a working prototype in order to answer some version of the question \textit{``Does it work?''} This is usually, though not always, done through evaluation with users. This process implicitly rests on an assumption that has held for decades: building a working interactive system is hard and slow, so a single, well-realized prototype plus a confirmatory evaluation of it constitutes a real advance.

But does this assumption still hold in 2026? Undergraduates now routinely walk into meetings with prototypes that would have consumed a grad student's whole semester a few years ago, put together with a coding agent in a few days. When time-to-working-implementation collapses, what should take its place? Economists distinguish between the intensive margin --- doing more of existing activities, and the extensive margin --- engaging in new activities.
Increasing the intensive margin of systems research by producing more of the same, just faster, is a recipe for exhausting  everyone: authors, reviewers, and readers. Instead, I think we should raise our expectations of what a UIST paper should accomplish and expand our extensive margin.

Deeper value will be found by adapting research methods around the new abundance of system prototypes and by thinking carefully about the promises and perils of this new reality. This argument isn't entirely novel: using generative AI only to construct the same prototype faster is analogous to replacing a factory’s steam engine with an electric motor while retaining its shafts, belts, and floor plan, when the real improvements come from redesigning the entire manufacturing process~\cite{david1990dynamo}. 
This argument also isn't true of all systems research. Some novel hardware, for example for human-computer integration~\cite{10.1145/3746058.3762829} or morphing interfaces~\cite{yu-morphingskin-uist2025}, remains hard. But for a growing share of the software systems we build to demonstrate an idea, \textit{``Can we build it and does it work as expected?''} is no longer as interesting a question as it used to be. 

Our community has discussed appropriate research methodologies for UIST before. Seminal papers pointed out that a naive, formulaic adherence to usability testing is not appropriate~\cite{olsen-evaluating-uist2007,greenberg-harmful}, and new technologies have led to new research methods before, e.g., through crowdsourcing~\cite{reinecke-labinthewild-2015}, or simulating users with~\cite{park-simulacra} or without~\cite{shi2024crtypist} LLMs. So, what methodological turn might we expect from ubiquitous code generation and fast system implementation?

\section{Three Speculative Directions}
Here are three speculative directions. Taken together, they point towards ways of asking different questions and drawing different conclusions rather than just accelerating what we already do. 

\subsection{Think About Multiples}
\begin{figure}[t]
  \centering
 \includegraphics[width=0.9\linewidth]{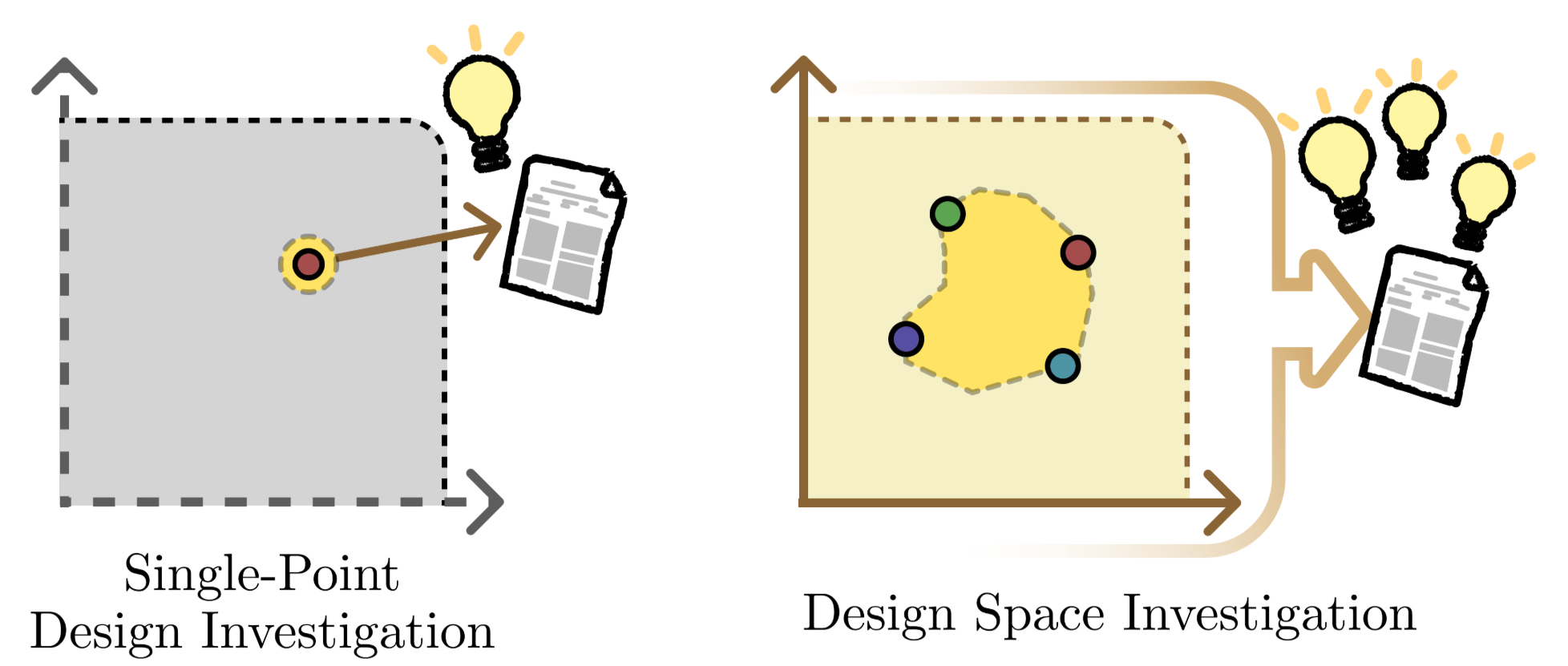}
  \caption{Left: Building and evaluating a single prototype tells us about its performance, but leaves larger questions about other potential solutions unanswered. Right: Evaluating multiple designs can answer such questions.}
  \Description{Conceptual representation of two design spaces, one with a single point design, one with a family of designs giving rise to subspace of feasible solutions.}
  \label{fig:design-space}
\end{figure}

Where it was once feasible to build exactly one system, as an end result of many sequential design decisions, we can now build entire families of prototypes and empirically map the tradeoffs between them. Design space analysis doesn't have to be a static diagram as a thought exercise in a paper. It can become a population of systems, built and compared to each other (see Figure~\ref{fig:design-space}). Researchers have already shown the benefits of parallel prototyping~\cite{dow-parallel-prototyping-2010},  evaluation of multiple alternatives~\cite{tohidi-design-2006,10.1145/3711838}, and structured thinking about design spaces~\cite{card-designspace-1991}.
Now is the time to take the lessons about the utility of designing with multiples to heart for our research process. Ideally, multiple prototypes can instantiate different theories, and thus contribute to resolving open questions in the literature. For example, Kazemitabaar et al. recently compared seven different cognitive engagement techniques in one paper~\cite{10.1145/3708359.3712104}, and Adar et al. showed that LLMs can regenerate (``revibe'') baseline comparisons from the past literature to better evaluate new interaction techniques~\cite{adar2026revibing}. Wherever automated evaluation is possible, we can now run sensitivity analyses of major decisions across whole families of designs, the way we sometimes already do for prompts or model choices, instead of defending the one configuration we happened to implement. 
However, if evaluation cannot be automated or accelerated, evaluation effort will become the new bottleneck that constrains how many ideas can be tested.
\begin{figure}[b]
  \centering
 \includegraphics[width=\linewidth]{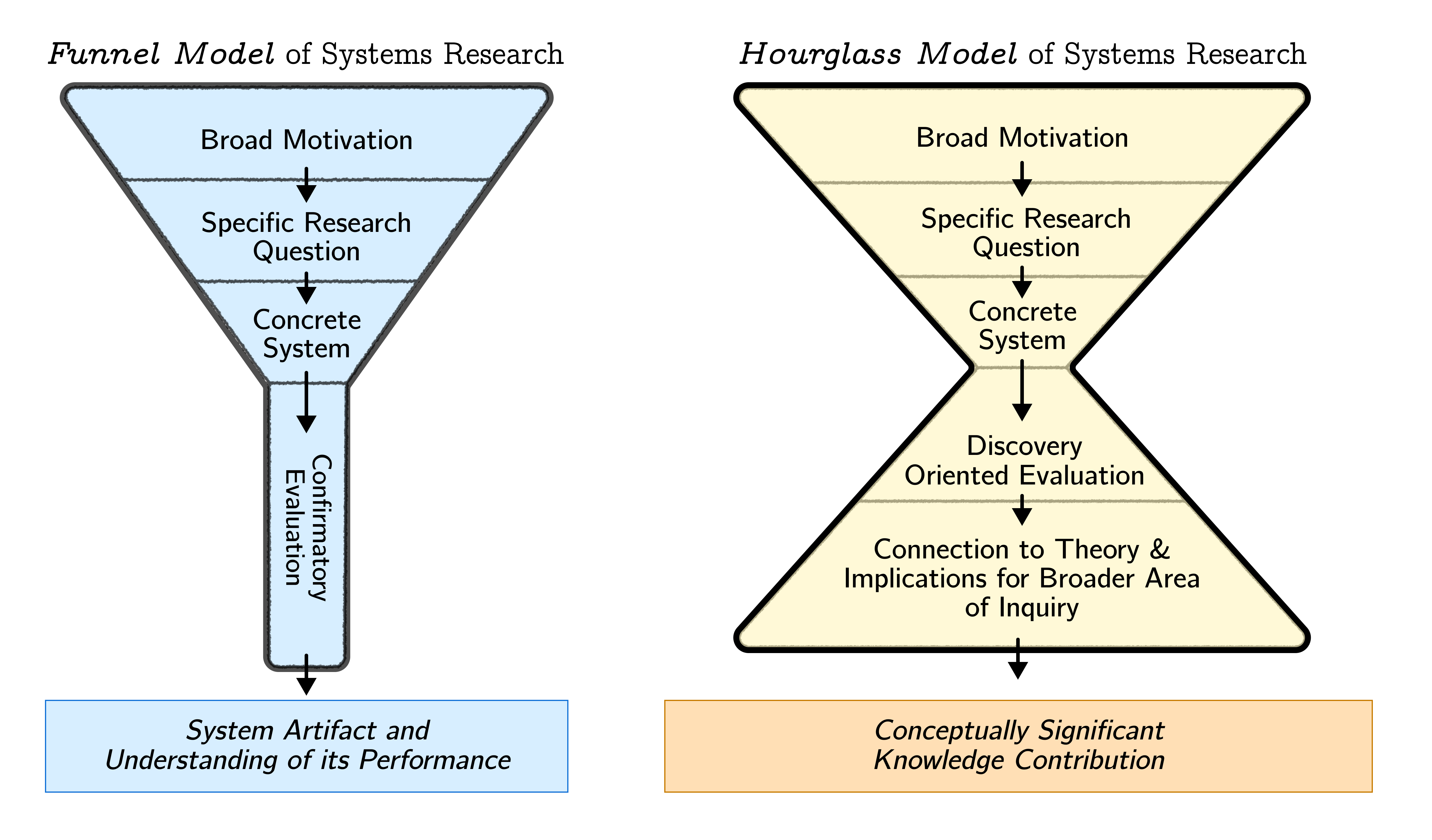}
  \caption{Confirmatory evaluations are an example of a ``funnel'' process. In contrast, discovery-oriented evaluations go beyond a narrow evaluation of the artifact towards broader implications, and represent an ``hourglass'' process.}
  \Description{Conceptual representation of two research processes, one shaped like a funnel and one shaped like an hourglass.} \label{fig:funnel_hourglass}
\end{figure}

\subsection{Ask More of the Evaluation}
Confirmatory evaluations can tell us a system works as intended, which is useful, but they rarely tell us much that generalizes beyond the specific prototype under study (Figure~\ref{fig:funnel_hourglass}, left). Other areas of HCI have long treated deployments as a generative method — a way to learn things the researchers hadn't already suspected, not just confirm the things they had. Historically, that kind of \emph{discovery-oriented evaluation} was often done with off-the-shelf technology, because building something substantively novel and hardening it to the point where it can be studied in the wild long enough to be surprised by its usage was too onerous. Now we can both build the novel system and use it as an artifact to generate knowledge that goes beyond affirming that ``it works'' (Figure~\ref{fig:funnel_hourglass}, right). Techniques include deploying systems as a probe~\cite{hutchinson_technology_2003} with a community to learn about``unknown unknowns'', developing strong concepts~\cite{hook:strongconcepts:10.1145/2362364.2362371}, or deploying systems in field experiments~\cite{Oulasvirta2009}. We can go beyond confirming our intuitions towards building deeper insights that arise when people encounter the cutting edge of what is technologically feasible.  As an example, almost no one remembers the system we built for ``Why Johnny Can't Prompt''~\cite{10.1145/3544548.3581388}, but the broader an\-alysis of prompting strategies that it enabled had an impact. One unique advantage for the UIST community is our experience with generating novel technologies that give rise to significantly different user experiences from the status quo, and thus carry potential for generating interesting new knowledge.

\begin{figure}[t]
  \centering
 \includegraphics[width=\linewidth]{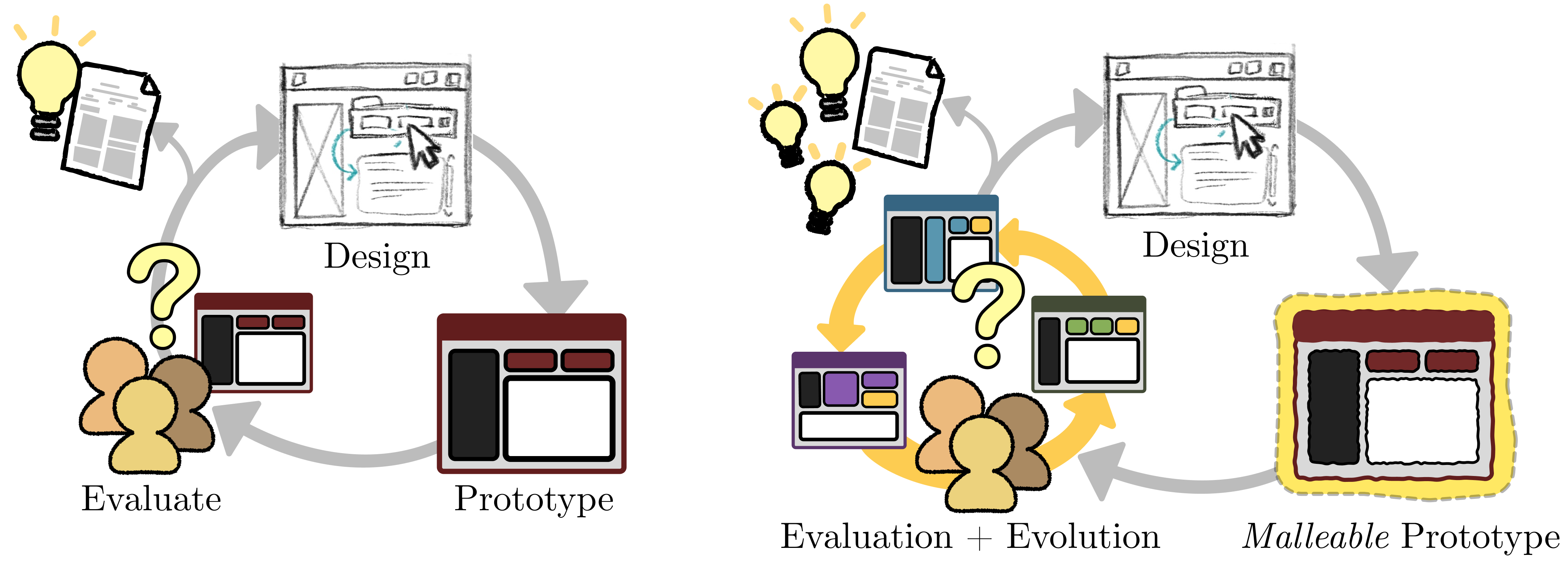}
  \caption{Left: Iterative design usually considers prototyping and evaluation as distinct stages. Right: Prototypes can continue to evolve during testing.}
  \Description{Conceptual representations of design processes. Left: Design-Prototype-Evaluate. Right: Design-Malleable Prototype-Evaluation+Evolution} \label{fig:evolution}
\end{figure}

\subsection{Develop New Methods for Software That Keeps Evolving}

While iterative design is widely taught and followed in HCI, we still often assume a hard boundary between prototyping and evaluation steps (Figure~\ref{fig:evolution}, left). Paper prototyping~\cite{snyder2003paper} and Wizard of Oz testing~\cite{Kelley1984WizardOfOz} were introduced precisely because real-time software iteration used to be unthinkable. Pen and paper, or human-produced responses were the only media fluid enough to change mid-study. Researchers also argued that low-fidelity prototypes are preferable in early design because they prevent premature fixation on irrelevant details~\cite{10.1145/223904.223910}. 
The prototyping landscape shifts when agents can re-write UIs and application code instantaneously. 

What would it look like to run a study where the artifact is constantly re-written during the session (Figure~\ref{fig:evolution}, right)? A participant's confusion can trigger a live regeneration of the interface they are using. Would this be a boon or a burden? While the literature already has examples of iteratively changing and deploying interfaces over months (e.g., ~\cite{10.1145/3411764.3445409}), the shift to evolving working interfaces in real time (which can also generate their own instrumentation) will likely require new methods. It raises questions about consent (researchers and participants won't know what will be generated ahead of time); analysis (how to compare participants if everyone ends up with a different interface); and replicability (the generated interfaces are never the same), among others.  More fundamentally, perhaps interfaces will not just change during a study, but will be generated on the fly during regular use as well~\cite{bernstein2026interface}. In that case, how do we evaluate the meta-properties of the interface generator instead of the produced UIs?

\section{What Will Your Future UIST Contribution Look Like?}
Take these as provocations to rethink what counts as a contribution once implementation is no longer the bottleneck. 
There is an opportunity to change the shape of our research process, and the format of our systems papers as a result. A paper that reports on multiple related prototypes that enact competing theories and analyzes their tradeoffs could yield more insights than a single polished interface and a lab study. So could a paper about a system that participants themselves collaboratively and continuously evolved across a multi-month deployment in the wild. Deciding on a path forward may require a higher-level reconsideration of what we want the goals of our research community to be, a process that is also underway in mathematics and other academic areas transformed by AI~\cite{tao2026mathematicsageai}. Let's think about how we can most meaningfully reallocate our research time and process to go “deeper, further, and higher” in developing  contributions and building a sustainable research community. 

\section{Acknowledgments}
While UIST Visions are single-author documents, these ideas are based on many discussions with the ``Systems and Situations'' group, including J.D. Zamfirescu-Pereira, Shm Garanganao Almeda, Max Kreminski, Eric Rawn, Jingyi Li, and James Smith. Thanks to Shm Garanganao Almeda for illustrative figures and to Antti Oulasvirta, Marti Hearst, and Mike Kuniavsky for constructive feedback.

\bibliographystyle{ACM-Reference-Format}
\bibliography{vision}

\end{document}